# Fast, broad-band magnetic resonance spectroscopy with diamond widefield relaxometry

C. Mignon[1,#], A. R. Ortiz Moreno[1,#], H. Shirzad[2], S. K. Padamati[1], V. Damle[1], Y. Ong[1]
R. Schirhagl[1,*], M. Chipaux[1,2,*]

# these authors contributed equally

1. Groningen University, University Medical Center Groningen, Antonius Deusinglaan 1, 9713 AW Groningen, the Netherlands
2. Institute of Physics, École Polytechnique Fédérale de Lausanne (EPFL), CH-1015 Lausanne, Switzerland

* Correspondance: romana.schirhagl@gmail.com mayeul.chipaux@epfl.ch

## Abstract

We present an alternative to conventional Electron Paramagnetic Resonance (EPR) spectroscopy equipment. Avoiding the use of bulky magnets and magnetron equipment, we use the photoluminescence of an ensemble of Nitrogen-Vacancy centers at the surface of a diamond. Monitoring their relaxation time (or T1), we detected their cross-relaxation with a compound of interest. In addition, the EPR spectra are encoded through a localized magnetic field gradient. While recording previous data took 12 minutes per data point with individual NV centers, we were able to reconstruct a full spectrum at once in 3 seconds, over a range from 3 to 11 Gauss. In terms of sensitivity, only $0.5\ \mu L$ of a $1\ \mu M$ hexaaquacopper (II) ion solution was necessary.

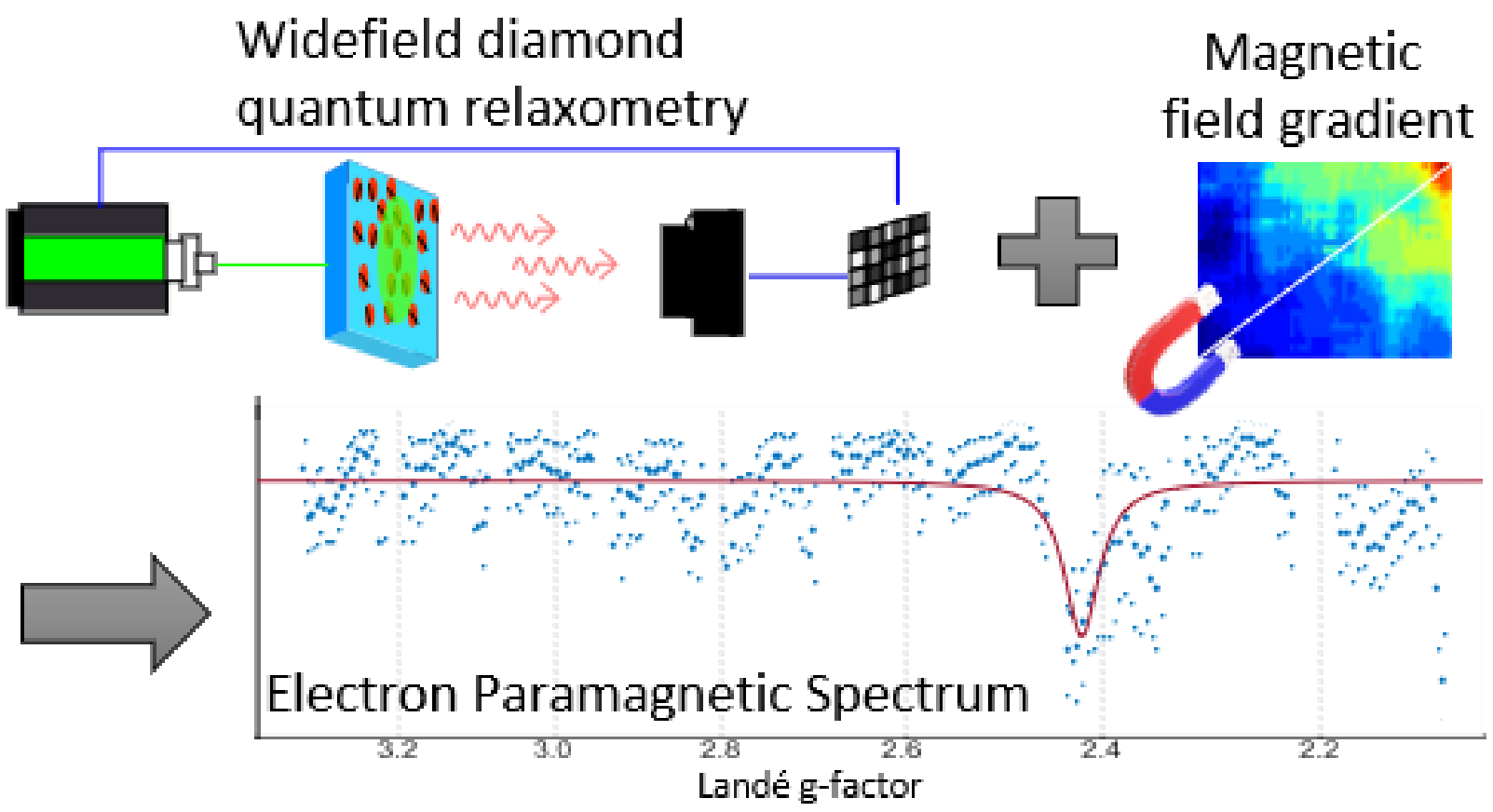

## Keywords

Electron-Paramagnetic Resonance (EPR) allows to detect and characterize chemicals with unpaired electrons [1]. Conventionally, a radiofrequency (RF) signal is absorbed by the compound of interest when it reaches its resonance. This conventional method however suffers from two intrinsic limitations. Due to the Boltzmann distribution, at room temperature, only very few spins are in position to absorb the RF. This drastically limits the sensitivity of the method [2] and necessitates either strong magnetic fields or low temperature. Besides, obtaining a full spectrum of the compound requires to scan over the resonances, either sweeping the magnetic field or the RF signal frequency.

A promising alternative is emerging from the diamond magnetometry field. It is based on the Nitrogen-Vacancy (NV) center, a point defect whose optical properties are sensitive to surrounding physical quantities. It is photoluminescent, both its ground and excited electronic states are spin triplets that can be initialized and readout by optical means [3]. These features called Optical Detection of Magnetic Resonance (ODMR) [4] allow to advantageously convert RF signals into optical ones which can be observed under standard optical microscopes [5]. While ODMR in general opened the way to detect individual electron spins in 1993 [6,7] NV centers represent a powerful alternative to standard EPR techniques. With this method detection of EPR spectra of different compounds [8,9,10] has been achieved with a sensitivity compatible with individual electron spins, external to the diamond [11]. Further it is possible to image EPR signals [12,13], or to perform spectroscopy [14,15]. This found various applications in chemistry, in monitoring reactions [16,17] and biology [18,19], in detecting free radicals in living cells [20,21].

Inherited from conventional magnetic resonance techniques, several measuring modes can be used when working with NV-centers. Different microwave and/or optical pulsing sequences to manipulate the NV-centers electronic state, can render the NV-center more sensitive or specific for spins with a certain Larmor frequency [22,23,24]. Among them, the so-called relaxometry measurements, or T1, is particularly useful. This sequence does not require microwave signals which can be challenging to use when imaging large surfaces [25]. In addition, microwaves are strongly absorbed by water. Both effects conflict with measuring in biological samples or solutions [26]. In T1 measurements, the NV center is pumped to its bright spin state $|m_S = 0 >$ with an optical pulse. After a variable “dark-time”, another optical pulse probes the decay of the NV centers to a darker thermal equilibrium of states $|m_S = 0 >$ $|m_S = +1 >$ and $|m_S = -1 >$. The resulting characteristic time of that dynamics, called relaxation time or T1 is shortened in presence of magnetic noise [27,16,17].

T1 cross-relaxation enables the retrieval of EPR spectroscopic data. It uses the fact that T1 is even further decreased when the NV center’s spin energy transition is at resonance with the one of the target compound. (Fig. 1D). As shown in [14] this allows to collect an entire spectrum of the P1 centers within the diamond. However, similarly as in conventional EPR measurements, the acquisition over the possible resonance position requires to scan the magnetic field. As a result, time resolution is limited. For instance, in [14], 12 minutes are required for each data point corresponding to 13 hours for a full spectrum over 12 Gauss.

In previous work [28] we used a magnetic field distribution (or gradient) over a bulk diamond filled with NV-centres to spread their resonances over space. This allowed to perform fast-high bandwidth spectroscopy of a microwave signal by imaging the NV center photoluminescence directly. This method has recently been further expanded up to a range of 25 GHz and a 40 dB dynamics [29]. Here we use a similar magnetic gradient to spread the NV center and target spin resonances in space (Figure 1 B,C). As a result, at a specific position along the gradient, the transition energy of the NV center matches the one of the target fulfilling and the cross-relaxation condition. At this position, the T1 is decreased (Figure 1 F). Each compound with a certain Landé “g-factor” would be in resonance with nearby NV centers at a specific position along the gradient. The EPR spectrum is then encoded along the gradient direction. The orthogonal coordinate is used for signal integration. This idea was patented [30] and is first demonstrated here.

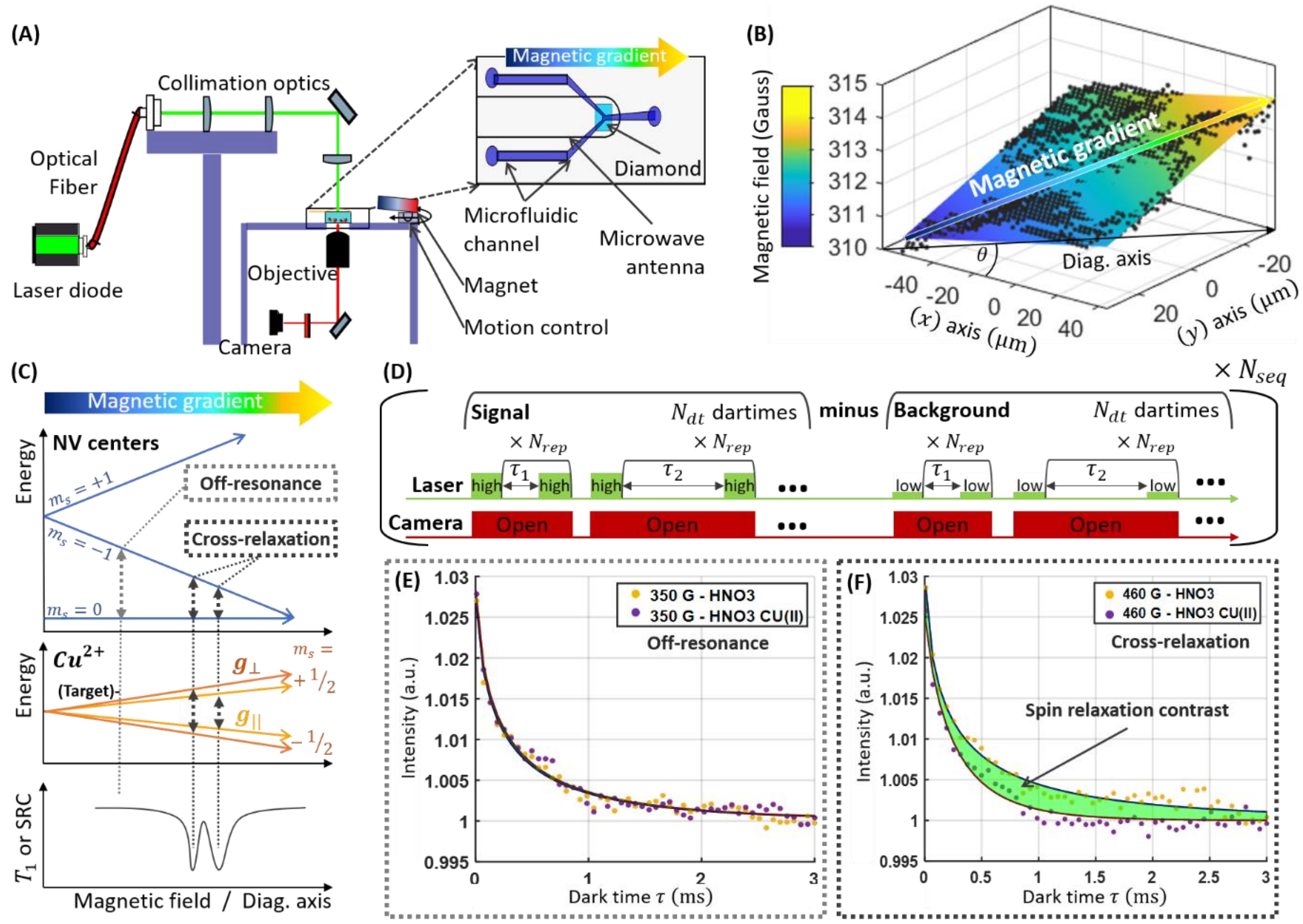

*Figure 1: Main principles. (A) A homemade fluorescence microscope images NV centers close to the surface of a bulk diamond. The sample is lighted by a laser diode driven by a current driver. Their fluorescence is collected by a microscope objective and imaged on a camera. The diamond is inserted in a microfluidic device for changing the fluid at its surface. A permanent magnet attached on a goniometer mount allows to generate a magnetic field distribution oriented along the NV center axis. A copper wire surrounds the diamond and is connected to a microwave generator. It acts as a microwave antenna for imaging the magnetic field to calibrate the instrument. (B) Typical magnetic field cartography obtained from the fluorescence image with an aligned magnet at 23 mm from the diamond. The coloured plane is adjusted from Eq. 4. The $(x)$ and $(y)$ axes on the diamond sample correspond to the rows and columns of the camera. The main direction of the magnetic gradient (Diag. axis) at an axis $\theta$ with respect to $(x)$ is also indicated in black. (D) Energy level diagrams of the NV centre ground state and a single spin compound of interest highlights the cross-relaxation condition. (C) T1-relaxation sequence. (E) and (F) relaxation curves computed from the difference of the signal (high current) to the background (low current). The yellow curve corresponds to the reference (nitric acid without copper sulphate), while the signal with 1 $\mu M$ copper sulphate is shown in purple. (E) Relaxometry curves at 350 G (Away from the cross-relaxation condition) and (F) at 460 G (at the cross-relaxation condition). The spin relaxation contrast corresponds to the area between the two curves. Detecting its response over a wide range of*

*magnetic fields allows a rapid screening of the cross-relaxation position of a certain compound of interest and compute an Electron Paramagnetic Resonance spectrum (see Fig.5 and 6).*

# Materials & Methods

## Wide-field microscope with quantum sensing capabilities

The initial calibration and relaxometry measurements were performed on a homemade wide-field microscope (Figure 1A). The excitation source is a 520 nm Dilas laser diode module (Coherent) modified to contain only 2 of the 3 internal laser diodes of the standard model. This allows to comply with the electrical characteristics of the Model 762 driver (Analog Modules) which directly modulates the current sent to the laser diode to produce the optical pulses. The driver is interfaced to the computer with an Aardvark I2C/SPI Host Adapter. The optical beam is sent through an optical fiber, collimated with an F230SMA integrated collimator (Thorlabs). The beam is further corrected by a couple of lenses and focused onto the diamond sample from its top with a wide-field lens (f=200 mm), delivering an effective power of 60 mW on the surface of the diamond.

The NV center fluorescence is collected by a 60X, 0.7 NA microscope objective (Olympus), selected by a 600 nm long-pass filter and focused by 175 mm focal field lens to the SCMOS sensor of a $2560 \times 2160$ pixels Zyla 5.5 camera (Andor). While the microscope images an area of approximately $280 \times 240\ \mu\mathrm{m}$, over the diamond, a subset is often chosen for the different acquisitions detailed hereafter.

The magnetic field comes from a permanent magnet (disk, 25 mm diameter, 7 mm thick, N42, supermagnet, S-25-07-N). For positioning the magnet, one large goniometer whose rotation axis coincides with the optical direction holds a radial translation stage and two tilt goniometers.

A microwave signal is generated by a generator HMC-T2220 (Hittite) and delivered to the diamond through a copper wire loop with a diameter of a few hundred microns and 3 cm length connected to a load resistance.

The instrument is controlled using a home-made LabVIEW program to automatically acquire T1-relaxation sequences and microwave resonance spectra. The data are processed with MATLAB.

## Diamond samples

A $2 \times 2 \times 0.5\ \mathrm{mm}^3$ electronic grade single crystal diamond plate was purchased from Element Six. To avoid spin-mixing and dramatic measurement contrast loss when the magnetic field increases, the NV centers need to be aligned with the magnetic field [31]. To ease the alignment on a wide-field surface, one of the diamond faces has been cut and polished along the [110] crystal orientation such that two out of the four crystallographic orientations are in that plane. The diamonds final size was $2 \times 2 \times 0.5\ \mathrm{mm}^3$ with a $1 \times 0.5\ \mathrm{mm}^2$ (110) surface available. The diamond was then cleaned using tri-acid (1:1:1 $HNO_3$:$HClO_4$:$H_2SO_4$) boiling under reflux for 1 h at around 300°C [32]. Then the [110] face was implanted with diatomic nitrogen with an energy of 12 KeV and a dose of $8.27 \times 10^{12}$ $cm^{-2}$.

These conditions create a layer of nitrogen and vacancies approximately 5 nm [33] below the diamond surface. After implantation the diamond was annealed under vacuum at 800°C for 3 hours [34] and tri-acid cleaned again. The NV centers on the surface have a $T_2^*$ of around 300 ns which is relatively stable on the entire range of applied magnetic fields (supplementary Figure 7s).

## EPR Sample preparation

As an EPR sample, we prepared a Copper II ions (a typical reference EPR sample already used in with NV centers [35]). We prepared a nitric acid solution by diluting concentrated nitric acid (90 %, Sigma-Aldrich, 695041) in MilliQ water. The sample itself was prepared by dissolving copper sulphate powder (Merck, 2790) in nitric acid solution to favor the formation of copper II hexaaqua $Cu(H2O)_6$ complex ions, which have a well-known EPR spectrum [36]. The final concentration of the solution is approximately $1\ \mu\mathrm{M}$. The sample was pumped through a Y shaped microfluidic channel (0.1 mm width, 0.5 mm thick) on a PDMS chip with a syringe.

## Magnetic field Imaging

### Acquisition

To image the magnetic field distribution, we selected a microwave frequency range based on our prior knowledge on the magnetic field, and an area of interest of $220 \times 130\ \mu\mathrm{m}$ with a $2 \times 2$ initial binning (images of $1000 \times 600\ pixels$) where the photoluminescence intensity is sufficient. A pair of images was acquired at each frequency step, one with microwaves, the other without. The relative photoluminescence loss between the two corresponds to the ODMR response at this step. Similarly, as in [37], the whole frequency sweep results in a 3-dimensional volume of data, where the two first directions are the spatial coordinates of the photoluminescent image and the third represents the frequency. As a consequence, each pixel possesses the ODMR responses at all frequencies.

### Processing and visualisation

Before any visualization or processing, the images are further binned down to $100 \times 60$ pixels. At first, one can intuitively visualize the gradient by imaging slices of the acquired volumes of data. With prior knowledge of its direction according to the diamond and setup geometry. Figure 3 and supplementary S3 represent slices (one-pixel width) along a diagonal axis rotated by 30 degree with respect to the horizontal axis as define by the camera (*e.g.* along the magnetic gradient). The images show the photoluminescence response as a function of microwave frequency and the diagonal axis.

The images are then further cropped to an area with sufficient contrast. Each binned pixel was then fitted by a Lorentzian function described by Eq. 1. (Figure 4A and supplementary 4S).

$$Lor(f) = \mathrm{A}\left(1 - \frac{C}{1+\left(\frac{f-f_0}{w}\right)^2}\right) \tag{1}$$

This was achieved by adjusting A (supposedly close to 1) the contrast C, the linewidth $w$ and the line position $f_0$ . Finally, we applied a $9 \times 9$ pixel median filter to remove pixels where the fit got “lost”. The measured magnetic field $B_{\parallel}$ at each pixel position is then given by the Zeeman effect on the NV center.

$$B_{exp} = (D_{NV} - f_0)/\gamma_{NV} \tag{2}$$

where $D_{NV} = 2.87\ GHz$ is the NV center zero-field splitting and $\gamma_{NV} = 2.80\ \mathrm{MHz}\cdot\mathrm{G}^{-1}$ its gyromagnetic ratio.

## Spin relaxation imaging

### Acquisition

Our T1-relaxation sequence differs from most diamond magnetometers found in the literature. This is due to the use of our specific directly electrically modulated laser module as excitation source. This solution appeared cheaper and easily implemented. Unfortunately, we measured that it reaches up to a maximum of 100 extinction ratio between the on and off status. Therefore, each of the $N_{dt}$ darktimes are repeated $N_{rep}$ times while the camera remains open. Both the relevant signal during the laser pulse and the background of the “off times” are acquired. We therefore acquire sequentially the NV center image response to either a high-current density (laser on) and a low-current density (laser nearly off). The numerical difference between the images suppresses the background and leads to our T1-relaxation curves.

$N_{rep}$ being kept, the exposure time therefore depends on $\tau$. The background is suppressed by an “off acquisition” with the same darktimes and $N_{rep}$ repetitions. The order of the dark times was randomized. The whole sequence including all the darktimes was repeated $N_{seq}$ times, changing the randomization. Each 60 mW laser pulse lasted $10\ \mu\mathrm{s}$. $N_{dt} = 50$ darktimes were linearly spaced from $10\ \mu\mathrm{s}$ to $3\ \mathrm{ms}$, with $N_{rep} = 2500$ and $N_{sec} = 10$. The area of interest (AOI) of the acquired data corresponds to $180 \times 90\ \mu\mathrm{m}^2$ on the diamond corresponding to $800 \times 400$ pixels with a $2 \times 2$ binning. In a similar manner, we obtain 3-dimensional data in which each pixel of the image is a T1 relaxation curve.

For the faster acquisitions (Figure 6), the process remains identical except that an AOI of $95 \times 95\ \mu\mathrm{m}$ was chosen corresponding to $170 \times 170$ pixels binned by $5 \times 5$. Further, only two dark times $\tau_s = 10\ \mu\mathrm{s}$ and $\tau_l = 3\ \mathrm{ms}$ were acquired with $N_{rep} = 1000$ and $N_{sec} = 1$.

### Processing and visualisation

Instead of fitting each relaxation curve, a Spin Relaxation Contrast (SRC) inspired from [38] was obtained as follows. Each relaxation curve was first normalized by its tail (averaged over the five last datapoints). The SRC is then the area between the curves for the solution of interest, and the one for the control (Figure 1E,F). For the simpler case where only two dark times are acquired, this leads to:

$$SRC_{x,y} = \left(1 + \left(\frac{PL_{x,y}(\tau_s)}{PL_{x,y}(\tau_l)}\right)\right)_{ref} - \left(1 + \left(\frac{PL_{x,y}(\tau_s)}{PL_{x,y}(\tau_l)}\right)\right)_{Cu^{2+}} \tag{3}$$

Where $PL_{x,y}$ corresponds to the photoluminescence acquired on the camera corresponding to the position (x,y) on the diamond, excluding ($ref$) or including the $Cu^{2+}$ ions.

# Results

The idea behind our instrument is to identify the cross-relaxation condition along one spatial direction on the diamond (Figure. 1C) and attribute the g-factor of the compound which engendered it (Figure 5B). For compounds with a g-factor approaching 2, the NV centers need to be placed in a magnetic configuration close the ESLAC [14] around 500 G. This requires a good aligning of the magnetic field to one of the NV centres directions to avoid any spin mixing and loss in ODMR contrast [31]. This magnetic distribution then needs to be characterized before acquiring relaxometry data.

## Device pre-configuration

### Magnetic field aligning procedure

The degrees of freedom for the magnet include: height and distance with respect to the diamond, tilt angles and one rotational angle in the plane of the diamond surface (goniometer stage). The last which defines the angle between the axis of the magnet and the normal edges of the diamond surface is the most significant.

A first rough alignment was performed at a magnetic field of about 100 G, e.g. with a magnet at 35 mm from the diamond. To this end, we monitored the ODMR spectra averaged on the central pixels at the AOI while rotating the magnet along the plane (Figure 2A). Among the four crystallographic orientations the NV center can take two of them are in the (110) plane of the active surface. We measured the evolution of the 4 |0> to |-1> transitions related to the projection of the magnetic field on the considered NV center axis (Figure 2B). For an angle around 30-35 degrees from the edge axes of the diamond, the transition of lowest frequency (label NV1) is at 2.55 GHz at its minimum. The three others (label NV 2-3-4) nearly join at 2.8 GHz. This indicates that the NV centers along axis 1 are roughly aligned with the magnetic field while by symmetry, the three others are submitted to similar projections. The obtained angle corresponds to the one between the [111] targeted axis and the edges of the (100) diamond main face.

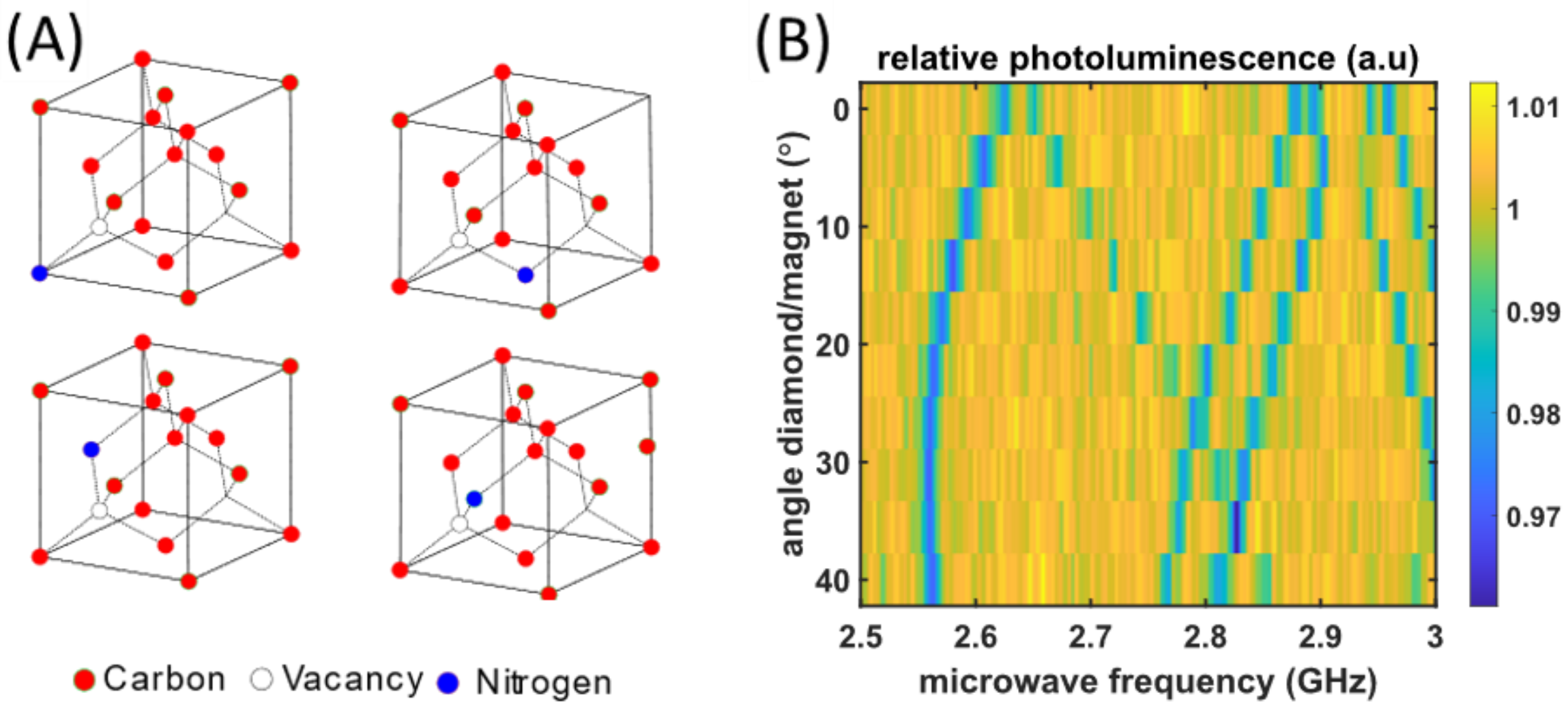

*Figure 2: Alignment of the magnetic field along one of the NV center axes. (A) NV centers along the four possible crystallographic orientations (B) Microwave response spectra obtained when scanning the angle between the direction of the magnetic field and the normal axes of the diamond. We note the presence of 4 |0> to |-1> resonances between 2.5 and 2.87 GHz, and 2 of the symmetric |0> to |+1> resonances above.*

The spin mixing first make the ODMR line related to direction 2, 3 and 4 to vanish such that we cannot reconstruct the vector magnetic field with the four signals such as in [37]. When approaching the ESLAC (B>400 Gauss), the alignment of the NV axis 1 with the magnetic field becomes critical as well such that a slight mis-alignment induces a drop of the contrast [31]. This may allow for vector magnetometry [39]. In our case, while at low magnetic field (<100 Gauss), an alignment accuracy of +- 1 degree was required, at high magnetic field (>400 Gauss), the magnetic field needs to be aligned up to +-1' arcminute (1/60 degree) which is the goniometer resolution. To finalize the aligning, we therefore looked for ODMR contrast in three separate sections. In Figure 3, we can distinguish one unambiguous ODMR line in only in one (0°, -1'), two (+2', -2', -6) or all three selected sections of the surface (-4'). In that last case, we can consider that the magnetic field is aligned within a few arcmin resolution all over the AOI. The Eq 2 is therefore valid to calculate the magnetic field from the ODMR lines' position.

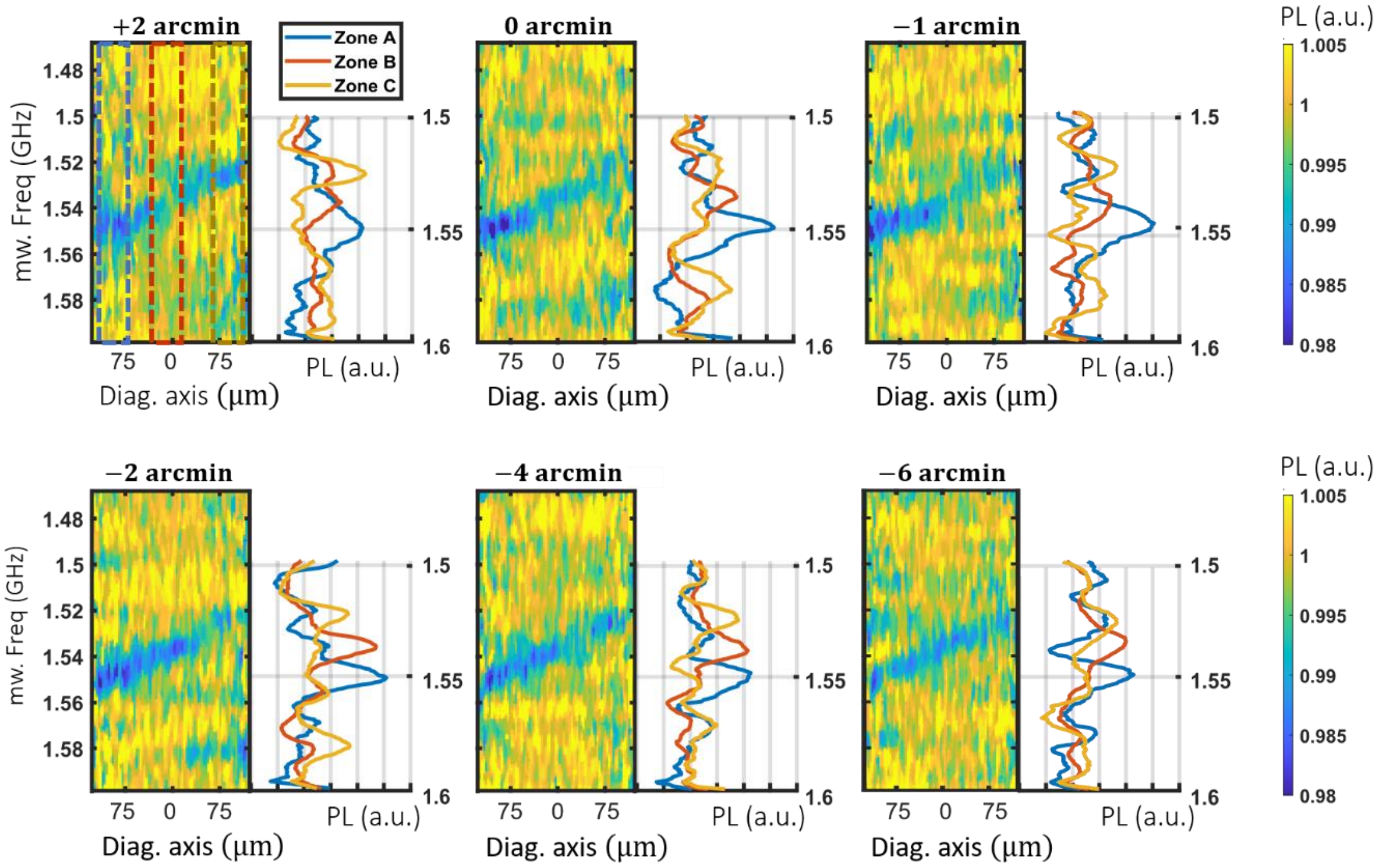

*Figure 3: Fine alignment of the magnetic gradient. Image of the photoluminescence response to microwave signals as a function of the frequency and the diagonal axis u, oriented along the magnetic gradient as a function of the tilt angle. The photoluminescence averaged over the blue, red and yellow regions of the first image are shown on the left.*

## Magnetic gradient characterization and field distribution imaging

As the translation axis of the magnet is not exactly radial to the diamond surface, the fine aligning might be lost when moving the magnet to low magnetic fields. However, the requirements are then sufficiently low to avoid the need of realigning it. The fitted magnetic field images are displayed in Figure 4A for certain relevant positions $z_m$ of the magnet (all of them are shown in supplementary Figure S4). A magnetic field distribution corresponding to a uniform gradient is adjusted over the $(x, y)$ plane of the image according to Eq. 4.

$$B_1(x, y) = B_0 + \nabla B \times (\cos(\theta)\, x + \sin(\theta)\, y) \tag{4}$$

where $B_0$ is the magnetic field value at the image origin considered at its center, $\nabla B$ its gradient and and $\theta$ its direction with respect to $x$. Approximating those parameters by an order 2 polynomial for $B_0$ and a linear regression for $\nabla B$ and $\theta$ (see Figure 4B, circled dots and Supplementary figure S1), we obtain a modelled magnetic field $B_{mod}(x, y, z_m)$ for every position $(x, y)$ on the camera and $z_m$ of the magnet, even where the magnetic field has not been measured.

$$B_{mod}(x, y, z_m) = B_0(z_m) + \nabla B(z_m) \times \big(\cos\big(\theta(z_m)\big) \cdot x + \sin\big(\theta(z_m)\big) \cdot y\big) \tag{5}$$

One can note that the fitted parameters $\nabla B$ and $\theta$ may differ from the regressions, caused by the difficult reconstruction of the magnetic field near the ESLAC. Not to be confounded with the angle between magnetic field $\theta(z_m)$, corresponds to the direction of the magnetic field gradient, known to be along the NV axis #1. It is therefore related to the position of the diamond with respect to the camera. Independent from the magnet position, it can be measured when the magnet is at longer distances. Including most of the data points (circled in Figure 4B), the Gradient $\nabla B(z_m)$ is well approximated by the linear regression. It also corresponds perfectly to the magnetic field evolution $B_0(z_m)$ considering the magnification of the microscope (Figure 4 B, Cyan line) and remains in accordance with the magnet model shown in Figure S1. In the following we therefore use the Eq. 5 with the parameters $B_0(z_m)$, $\nabla B(z_m)$ and $\theta(z_m)$ as define in Figure 4B.

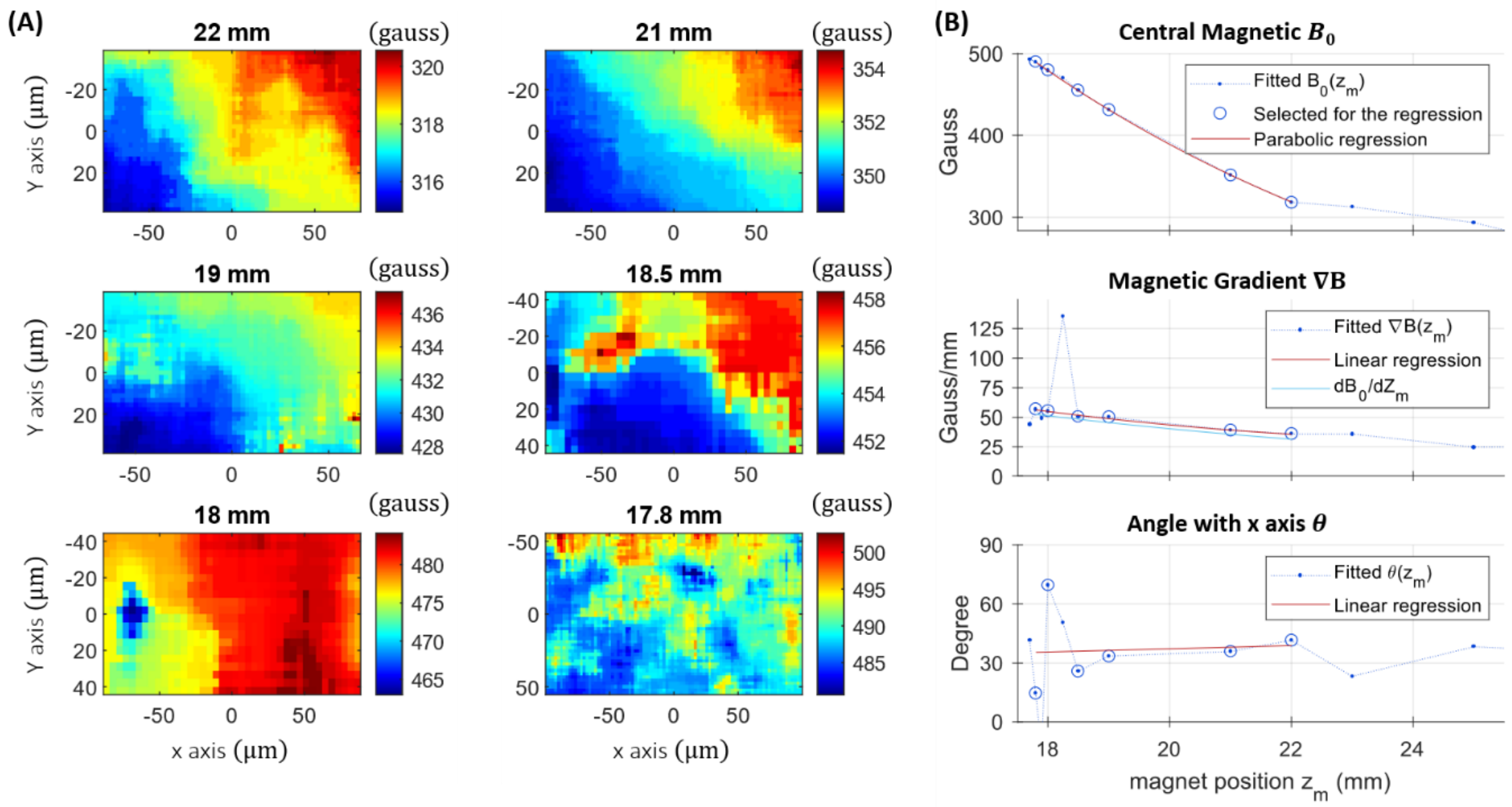

*Figure 4: Magnetic gradient visualization and characterization. (A) Magnetic field measured at different positions of the magnet based on Eq. 2. (B) Fitted parameters according Eq. 4 as a function for the magnet position, approximated by linear and parabolic regression (Eq. 5). The cyan line corresponds to the derivation of $B_0(z_m)$ with the magnet position $z_m$ considering the magnification of the microscope.*

## Microwave-free electron paramagnetic resonance spectroscopy

Once we successfully aligned the magnetic field gradient on the NV surface, we tested our system by acquiring the T1-relaxation signal in presence of copper (II) ions in solution. This involved a solution of nitric acid (1 mM) serving as blank and a solution containing the target, $1\ \mu\text{M}$ copper sulphate dissolved in the same nitric acid solution.

We measured T1-relaxation signals in various positions of the magnet from [21-17.7] mm, corresponding to a magnetic field range of [350-500] G. Averaging the signal over the entire AOI, we observe a shortening of the T1-relaxation at high magnetic field (>400 G) and a maximum around 460 G (Figure 5S). Instead of fitting a relaxation curve to all the pixels of the relaxation images, we calculated a Spin Relaxation Contrast (SRC) where both the blank and target solution were involved (See method Eq. 3). Figure 5A displays the SRC maps, for the different magnet positions. They are rotated by $\theta(z_m)$ degrees such that the vertical axis corresponds to a constant magnetic field, allowing to read the EPR spectrum horizontally. On SRC maps, the darker (bluer) areas corresponds to a shortened T1 when Cu II ions are introduced. More quantitatively, Eq. 5 allows to subsample each SRC map into an arbitrary number $N_{bin} = 10$, equal to the number of independent channels (see discussion section) of equally spaced magnetic bins. Their boundaries are the diagonals perpendicular to the gradient as sketched in Figure 6. The corresponding pixels of the SRC images are then grouped for statistic evaluation. The EPR spectrum in Figure 5B is obtained out of the median computed over each bin. The component g-factor can then be deduced from cross-relaxation resonance $B_{res}$ according to Eq. 6.

$$g_{Cu} = g_{NV}\left(\frac{D_{NV}}{\gamma_{NV}B_{res}} - 1\right) \tag{6}$$

where $g_{NV}$ is the effective g-factor of the NV spin 2.00287 (17) and $\gamma_{NV} = \frac{g_{NV}\mu_b}{h} = 2.80\ \text{MHz/G}$ is the NV center gyromagnetic ratio. Thus, the obtained resonance magnetic field value (463. +- 0.3 G) yields a g factor for copper of around 2.43 +- 0.03.

Copper II ions in solution (dissolved in water in acidic conditions) measured using standard EPR technology show two g-factors of $g_{\parallel}^{\text{Cu}} = 2.400$ and $g_{\perp}^{\text{Cu}} = 2.099$ (15). Typically, using standard EPR technology, in a solution at ambient temperature, a motion average g factor is expected to appear around 2.23 (484 G) for copper II ions (1). Yet, diamond-based EPR is not measuring the signal from a solution but rather from a thin layer of adsorbed ions on of the diamond surface. As developed in [21] the reason is that the sensitivity of NV centers to free radicals decreases very fast with their distance. Thus, in our spectrum, we assign the resonance peak (at $g = 2.43$ and 2.37) to the transition with $g_{\parallel}^{\text{Cu}}$ in an adsorbed and partially free configuration. The $g_{\perp}^{Cu}$ transition may also appear at the right limit of the spectrum at $495\ G$ and $500$ G corresponding to the range limit allowed by our magnetic field aligning.

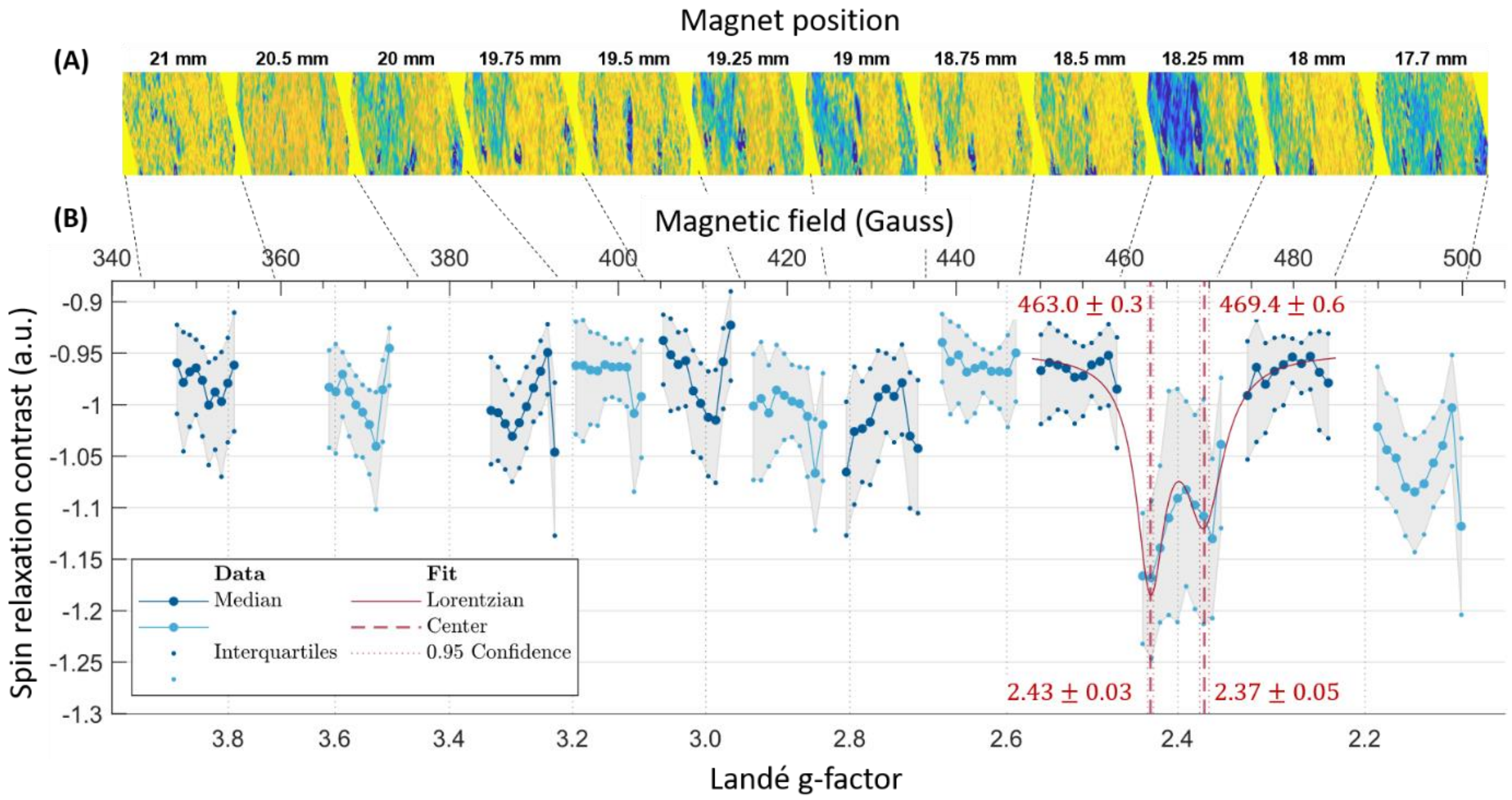

*Figure 5: EPR spectrum of copper sulphate in nitric acid solution. (A) SRC images at different magnet positions, rotated by $\theta(z_z)$ so that the spectrum can be read horizontally, and cropped down to the central 100 rows. (B) Copper EPR spectra obtain by subsampling the SRC images as described in the text. The displayed data correspond to the median and interquartile range obtained from above SRC sampled as described in the text, normalized by their common noise (e.g.* the average of the all 12 magnet positions*). The Alternance between blue and cyan data markers allows to distinguish the different positions of the magnet.*

## Performances.

If we neglect the laser “on times” (10 µs) with respect to the darktimes, the total acquisition time $T_{ac}$ corresponds to the sum of all applied dark times.

$$T_{ac} = N_{Rep} \times N_{seq} \times T_\tau \quad (7)$$

Where $N_{rep} = 2500$ is the number of times each darktime is repeated in a row, $T_\tau = \sum_{i=1}^{N_{dt}} \tau_i = 75$ ms the sum of all the darktimes, and $N_{seq} = 10$ the number of times the full sequence is repeated. This adds up to 31 minutes for each position of the magnet. The time for moving the magnet, calibrating the magnetic field and for acquiring the blank spectra (which can be reused for other acquisitions) are not considered here.

We note however, that, from the $800 \times 400$ pixels of the SRC images, each magnetic field bin of the spectra groups at least a number of pixels $N_{pix}(bin)$ exceeding 1000. The standard error made on spin relaxation contrast shown for each bin is therefore much smaller than the displayed interquartile range, by approximately $\sqrt{N_{pix}(bin)}$ times). We therefore attribute the remaining observed fluctuations, of larger amplitude, to artefacts or signals of unknown origins. As shown if supplementary paragraph 5 Figure 6S, taking a small subset of the acquired data with a virtual acquisition time of $T_{ac} = 10$ s (for each position of the magnet) does not significantly deteriorate the signal.

To further test that the acquisition can be further accelerated, we specifically realigned the magnet where the Cu resonance is visible with an AOI of approximately $95 \times 95$ μm.
We acquired a magnetic image according to previous methods (Figure 6B) and approximated the magnetic field according to Eq. 4. We then acquired a T1 series with only two dark times $10$ μs and $3$ ms with a total acquisition time adding to 3s. The SRC map is shown Figure 6C. Taking $N_{bin} = 100$, the EPR spectrum of Figure 6D can be obtained by aggregating the spin relaxation contrast along the diagonal binning described previously.

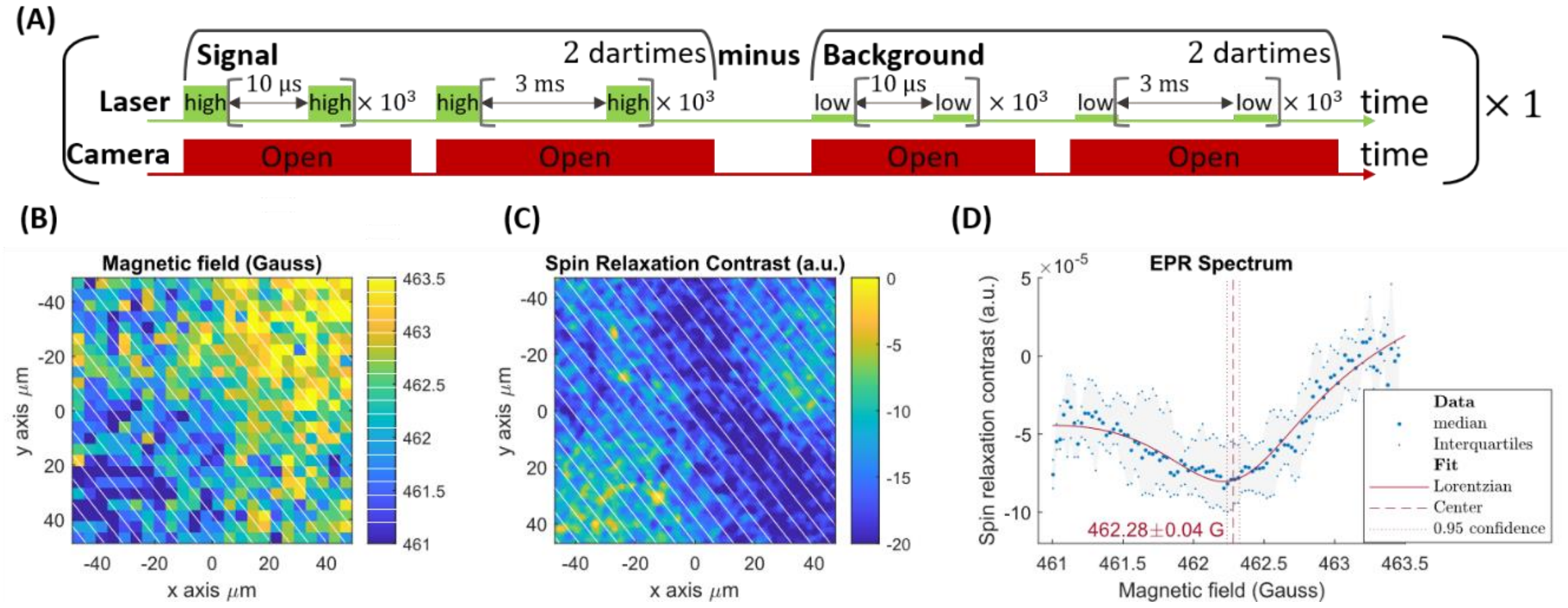

*Figure 6: 3 second acquisition EPR spectroscopy (Two darktimes). (A) Pulse sequence. (B) Magnetic field image of an approx.* $95 \times 95$ *μm area of interest. (B) SRC map over the same area. 25 out of the 100 magnetic field bins separation are shown in white in (A) and (B). The median and interquartile range of the SRC values of each beam allows to reconstruct the EPR spectrum in (C)*

# Discussion

We demonstrated microwave-free, scanning-free detection electron paramagnetic resonance spectroscopy at ambient temperature using diamond magnetometry. As a proof of principle, we retrieved the EPR spectrum of copper II ions. We obtained a magnetic resonance spectrum from a very low volume ($0.5$ μL) of a $1$ μM concentrated solution, where one spatial direction was reserved for the variation of magnetic field and the perpendicular direction was used for averaging. The relevant performance parameters are acquisition time, frequency (or g-factor) resolution and magnetic field range.
With respect to previous methods, including the one based on NV centers, the use of the gradient allows to acquire the spectra all at once, without a need to scan a magnetic field or a microwave frequency. While in [14], 12 seconds of acquisition time were necessary for each data point (13 hours for a 12 Gauss range spectra). The spectra shown in Figure 6 lasted 3 second only for a 5 times smaller range. Those times neither include the magnetic field calibration time nor the dedicated cleaning required to removed positively charged compounds.
With this protocol, the frequency resolution is in principle given by the heterogeneous linewidth of the NV center ${T_2^*}^{-1} \approx 3$ MHz. The linewith of the Cu(II) was larger than 3 MHz and thus not visible in our data. On that aspect, great recent progress has been made in creating high quality dense layers of shallow NV centers [40]. While the true coherence time T2 is heavily discussed in the published preprint [40], the information about the heterogeneous broadening is missing. Further improvement could be obtained from surface treatment to optimize the adsorption of the target spin to the surface. The surface chemistry of diamond is particularly rich and allows a variety of chemical reactions [41]. Finally, alternative ODMR defect such as in 2D materials [42] could further allow to decrease the distance between sensors and target spins and further increase the sensitivity.

The magnetic field range is approximately $11$ Gauss when using the larger AOI. It is limited by the total imaged magnetic field variation. Considering a linewidth of 3 MHz, for each position of the magnet, we obtain $\frac{range}{Linwidth} = 10$ independent spectral channels corresponding to 18 µm over the diamond, far above the optical resolution. This motivated us to still move the magnet to several positions to acquire a larger range. Besides, heterogeneity of the magnetic field directions over the imaged plane constituted our main limitation for contrasted measurement when approaching the ESLAC (512 G). This therefore also limits the maximum detection range of the instrument. In our case, each channel of the spectra is largely oversampled (with tens of thousands of pixels involved each time). Spreading the magnetic field more by using a stiffer gradient would constitute a major improvement. This however would require to design a specific gradient generator such that the field lines can be maintained as parallel as possible in the imaged plane, while allowing them to diverge in the orthogonal direction. The idea is that the magnetic field can vary significantly while staying aligned with the NV centers.
There is also room for improvement concerning the low extinction ratio of our laser pulsing system which induces a background to be removed and repumps the NV centre. This problem limits the T1 that can be observed and therefore the sensitivity of the NV centres.

Once the magnetic field distribution is known, the instrument does not require the use of microwaves to detect the electron paramagnetic resonance signal. This is an advantage over literature reports on detection of electron paramagnetic resonance in external targets using diamond magnetometry. The former work always relied on a reference control in presence of microwaves. Avoiding microwaves is beneficial for biological samples or samples in solution, which contain water and are thus very sensitive to microwave radiation.
Finally, we envision that the extra spatial direction, perpendicular to the magnet, could be used for other types of applications by not using a homogeneous sample but a sample which changes in space. This opens the way for deciphering reaction intermediates for instance by superimposing a microfluidic channel axis on the direction orthogonal to the magnetic field gradient.

# Acknowledgments

This project was financially supported by the Dutch research Council (NWO) for the Demonstrator grant (16088) and a VIDI grant (016.Vidi.189.002). MC acknowledge the support from the Swiss National Science Foundation under the ambizione PZ00P2_185824, VGD acknowledges the European Commission for a Marie Skłodowska-Curie scholarship (838494 - MagnetoVirology). The authors thank the ICT of the University of Groningen for support.

# Supporting Information Available:

The following files are available free of charge.

Supplementary document containing: a magnetic field simulation, conventional electron spin resonance spectroscopy data, measurements of the magnetic field for different magnet positions, Raw data for determining the spin relaxation contrast, reducing data acquisition with subsamples of the data, Ramsey fringes.

# References

[1] Prisner, T., Rohrer, M., & MacMillan, F. (2001). PULSED EPR SPECTROSCOPY: Biological. *Annu. Rev. Phys. Chem*, *52*, 279-313.

[2] Eaton, G. R., Eaton, S. S., & Rinard, G. A. (1998). *Frequency dependence of EPR sensitivity*. Wiley-VCH. Weinheim.

[3] Gruber, A., Dräbenstedt, A., Tietz, C., Fleury, L., Wrachtrup, J. and Von Borczyskowski, C., 1997. Scanning confocal optical microscopy and magnetic resonance on single defect centres. *Science*, *276*(5321), pp.2012-2014

[4] Suter, D. (2020). Optical detection of magnetic resonance. *Magnetic Resonance*, *1*(1), 115-139.

[5] Babashah, H., Shirzad, H., Losero, E., Goblot, V., Galland, C., & Chipaux, M. (2022). Optically detected magnetic resonance with an open source platform. *arXiv preprint arXiv:2205.00005*.

[6] J. Wrachtrup, C. v. Borczyskowski, J. Bernard, M. Orrit, and R. Brown, "Optically detected spin coherence of single molecules," Physical Review Letters **71**, 3565 (1993).

[7] Köhler, J. A. J. M. Disselhorst, M. C. J. M. Donckers, E. J. J. Groenen, J. Schmidt, andW. E. Moerner, "Magnetic resonance of a single molecular spin," *Nature* **363**:6426 363, 242–244 (1993).

[8] Alfasi, N., Masis, S., Shtempluck, O., & Buks, E. Detection of paramagnetic defects in diamond using off-resonance excitation of NV centres. *Physical Review B*, **99**(21), 214111 (2019)

[9] Wood, J.D., Tetienne, J.P., Broadway, D.A., Hfall, L.T., Simpson, D.A., Stacey, A. and Hollenberg, L.C., Microwave-free nuclear magnetic resonance at molecular scales. *Nature communications*, ***8*(1)**, 1-6 (2017).

[10] Hall, L.T., Kehayias, P., Simpson, D.A., Jarmola, A., Stacey, A., Budker, D. and Hollenberg, L.C.L., Detection of nanoscale electron spin resonance spectra demonstrated using nitrogen-vacancy centre probes in diamond. *Nature communications*, ***7*(1)**, 1-9 (2016).

[11] Grinolds, M.S., Hong, S., Maletinsky, P., Luan, L., Lukin, M.D., Walsworth, R.L. and Yacoby, A., Nanoscale magnetic imaging of a single electron spin under ambient conditions. *Nature Physics*, ***9*(4)**, 215-219 (2013).

[12] Steinert, S., Ziem, F., Hall, L.T., Zappe, A., Schweikert, M., Götz, N., Aird, A., Balasubramanian, G., Hollenberg, L. and Wrachtrup, J., Magnetic spin imaging under ambient conditions with sub-cellular resolution. *Nature communications*, ***4*(1)**, 1-6 (2013).

[13] Simpson, D.A., Ryan, R.G., Hall, L.T., Panchenko, E., Drew, S.C., Petrou, S., Donnelly, P.S., Mulvaney, P. and Hollenberg, L.C., Electron paramagnetic resonance microscopy using spins in diamond under ambient conditions. *Nature communications*, ***8*(1),** 1-8 (2017).

[14] Wood, J.D., Broadway, D.A., Hall, L.T., Stacey, A., Simpson, D.A., Tetienne, J.P. and Hollenberg, L.C., c. *Physical Review B*, ***94*(15)**, 155402 (2016).

[15] Purser, C.M., Bhallamudi, V.P., Wolfe, C.S., Yusuf, H., McCullian, B.A., Jayaprakash, C., Flatté, M.E. and Hammel, P.C., Broadband electron paramagnetic resonance spectroscopy in diverse field conditions using optically detected nitrogen-vacancy centres in diamond. *Journal of Physics D: Applied Physics*, ***52*(30)**, 305004 (2019).

[16] Perona Martínez, F., Nusantara, A.C., Chipaux, M., Padamati, S.K. and Schirhagl, R., Nanodiamond Relaxometry-Based Detection of Free-Radical Species When Produced in Chemical Reactions in Biologically Relevant Conditions. *ACS Sensors*. (2020).

[17] Barton, J., Gulka, M., Tarabek, J., Mindarava, Y., Wang, Z., Schimer, J., ... & Cigler, P. (2020). Nanoscale dynamic readout of a chemical redox process using radicals coupled with nitrogen-vacancy centres in nanodiamonds. *ACS nano*, *14*(10), 12938-12950.

[18] Chipaux, M., van der Laan, K.J., Hemelaar, S.R., Hasani, M., Zheng, T. and Schirhagl, R., 2018. Nanodiamonds and their applications in cells. *Small*, *14*(24), p.1704263.

[19] van der Laan, K., Hasani, M., Zheng, T. and Schirhagl, R., 2018. Nanodiamonds for in vivo applications. *Small*, *14*(19), p.1703838.

[20] Morita, A., Nusantara, A.C., Mzyk, A., Martinez, F.P.P., Hamoh, T., Damle, V.G., van der Laan, K.J., Sigaeva, A., Vedelaar, T., Chang, M., Chipaux, M. and Schirhagl, R., *(2022)* Detecting the metabolism of individual yeast mutant strain cells when aged, stressed or treated with antioxidants with diamond magnetometry. *Nano Today,* **48,** 101704.

[21] Sigaeva, A., Shirzad, H., Martinez, F. P., Nusantara, A. C., Mougios, N., Chipaux, M., & Schirhagl, R.. Diamond-Based Nanoscale Quantum Relaxometry for Sensing Free Radical Production in Cells. *Small*, 2105750 (2022).

[22] Loretz, M., Boss, J.M., Rosskopf, T., Mamin, H.J., Rugar, D. and Degen, C.L., Spurious harmonic response of multipulse quantum sensing sequences. *Physical Review X*, ***5*(2),** 021009 (2015).

[23] Wolf, T., Neumann, P., Nakamura, K., Sumiya, H., Ohshima, T., Isoya, J. and Wrachtrup, J., Subpicotesla diamond magnetometry. *Physical Review X*, ***5*(4)**, 041001 (2015).

[24] Barry, J.F., Schloss, J.M., Bauch, E., Turner, M.J., Hart, C.A., Pham, L.M. and Walsworth, R.L., Sensitivity optimization for NV-diamond magnetometry. *Reviews of Modern Physics*, ***92*(1)**, 015004. (2020).

[25] Opaluch, O. R., Oshnik, N., Nelz, R., & Neu, E. (2021). Optimized planar microwave antenna for nitrogen vacancy centre based sensing applications. *Nanomaterials*, **11**(8), 2108.

[26] Nesmelov, Y.E., Gopinath, A. and Thomas, D.D., Aqueous sample in an EPR cavity: sensitivity considerations. Journal of Magnetic Resonance, ***167*(1),** 138-146 (2004).

[27] Tetienne, J. P., Hingant, T., Rondin, L., Cavaillès, A., Mayer, L., Dantelle, G., ... & Jacques, V. Spin relaxometry of single nitrogen-vacancy defects in diamond nanocrystals for magnetic noise sensing. *Physical Review B*, **87**(23), 235436 (2013).

[28] Chipaux, M., Toraille, L., Larat, C., Morvan, L., Pezzagna, S., Meijer, J. and Debuisschert, T., 2015. Wide bandwidth instantaneous radio frequency spectrum analyzer based on nitrogen vacancy centres in diamond. *Applied Physics Letters*, *107*(23), p.233502.

[29] Magaletti, S., Mayer, L., Roch, J. F., & Debuisschert, T. A quantum radio frequency signal analyzer based on nitrogen vacancy centers in diamond. Communications Engineering, 1(1), 1-8. (2022)

[30] Chipaux, M; Schirhagl R Instantaneous magnetic resonance spectroscopy of a sample Patent: WO2018128543A1 (2018)

[31] Tetienne, J.P., Rondin, L., Spinicelli, P., Chipaux, M., Debuisschert, T., Roch, J.F. and Jacques, V., 2012. Magnetic-field-dependent photodynamics of single NV defects in diamond: an application to qualitative all-optical magnetic imaging. *New Journal of Physics*, *14*(10), p.103033.

[32] Brown, K. J., Chartier, E., Sweet, E. M., Hopper, D. A., & Bassett, L. C. (2019). Cleaning diamond surfaces using boiling acid treatment in a standard laboratory chemical hood. Journal of Chemical Health & Safety, 26(6), 40-44.

[33] Pezzagna, S., Naydenov, B., Jelezko, F., Wrachtrup, J., & Meijer, J. (2010). Creation efficiency of nitrogen-vacancy centres in diamond. New Journal of Physics, 12(6), 065017.

[34] Tetienne, J. P., De Gille, R. W., Broadway, D. A., Teraji, T., Lillie, S. E., McCoey, J. M., ... & Hollenberg, L. C. L.. Spin properties of dense near-surface ensembles of nitrogen-vacancy centres in diamond. Physical Review B, 97(8), 085402 (2018).

[35] Simpson, D.A., Ryan, R.G., Hall, L.T., Panchenko, E., Drew, S.C., Petrou, S., Donnelly, P.S., Mulvaney, P. and Hollenberg, L.C., Electron paramagnetic resonance microscopy using spins in diamond under ambient conditions. *Nature communications*, ***8*(1),** 1-8 (2017).

[36] Hoffmann, S. K., Goslar, J., Ratajczak, I., & Mazela, B. (2008). Fixation of copper-protein formulation in wood: Part 2. Molecular mechanism of fixation of copper (II) in cellulose, lignin and wood studied by EPR.

[37] Chipaux, M., Tallaire, A., Achard, J., Pezzagna, S., Meijer, J., Jacques, V., ... & Debuisschert, T. (2015). Magnetic imaging with an ensemble of nitrogen-vacancy centres in diamond. The European Physical Journal D, 69(7), 1-10.
[38] Simpson, D. A., Tetienne, J. P., McCoey, J. M., Ganesan, K., Hall, L. T., Petrou, S., ... & Hollenberg, L. C. (2016). Magneto-optical imaging of thin magnetic films using spins in diamond. Scientific reports, 6(1), 1-8.
[39] Zheng, H., Sun, Z., Chatzidrosos, G., Zhang, C., Nakamura, K., Sumiya, H., ... & Budker, D. (2020). Microwave-free vector magnetometry with nitrogen-vacancy centers along a single axis in diamond. Physical Review Applied, 13(4), 044023.
[40] Healey, A. J., Scholten, S. C., Nadarajah, A., Singh, P., Dontschuk, N., Hollenberg, L. C. L., ... & Tetienne, J. P. (2022). On the creation of near-surface nitrogen-vacancy centre ensembles by implantation of type Ib diamond. arXiv preprint arXiv:2210.16469.
[41] Nagl, A., Hemelaar, S. R., & Schirhagl, R. (2015). Improving surface and defect center chemistry of fluorescent nanodiamonds for imaging purposes—a review. Analytical and bioanalytical chemistry, 407, 7521-7536.
[42] Robertson, I. O., Scholten, S. C., Singh, P., Healey, A. J., Meneses, F., Reineck, P., ... & Tetienne, J. P. (2023). Detection of paramagnetic spins with an ultrathin van der Waals quantum sensor. arXiv preprint arXiv:2302.10560.

**Supplementary Information:**

# Fast, broad-band magnetic resonance spectroscopy with diamond widefield relaxometry

C. Mignon[1,#], A. R. Ortiz Moreno[1,#], H. Shirzad[2], S. K. Padamati[1], V. Damle[1], Y. Ong[1] R. Schirhagl[1,*], M. Chipaux[1,2,*]

# these authors contributed equally

3. Groningen University, University Medical Center Groningen, Antonius Deusinglaan 1, 9713 AW Groningen, the Netherlands
4. Institute of Physics, École Polytechnique Fédérale de Lausanne (EPFL), CH-1015 Lausanne, Switzerland
5. Zernike Institute for Advanced Materials, Nijenborgh 4, 9747 AG, Groningen, the Netherlands

* Correspondance: romana.schirhagl@gmail.com mayeul.chipaux@epfl.ch

# 1. Magnetic field simulation

In order to optimize the position of the cylindrical magnet, we performed a simulation using the expression derived in ([43]), and using the magnetization given by the dipole approximation as the magnetization in the axis of the magnet.

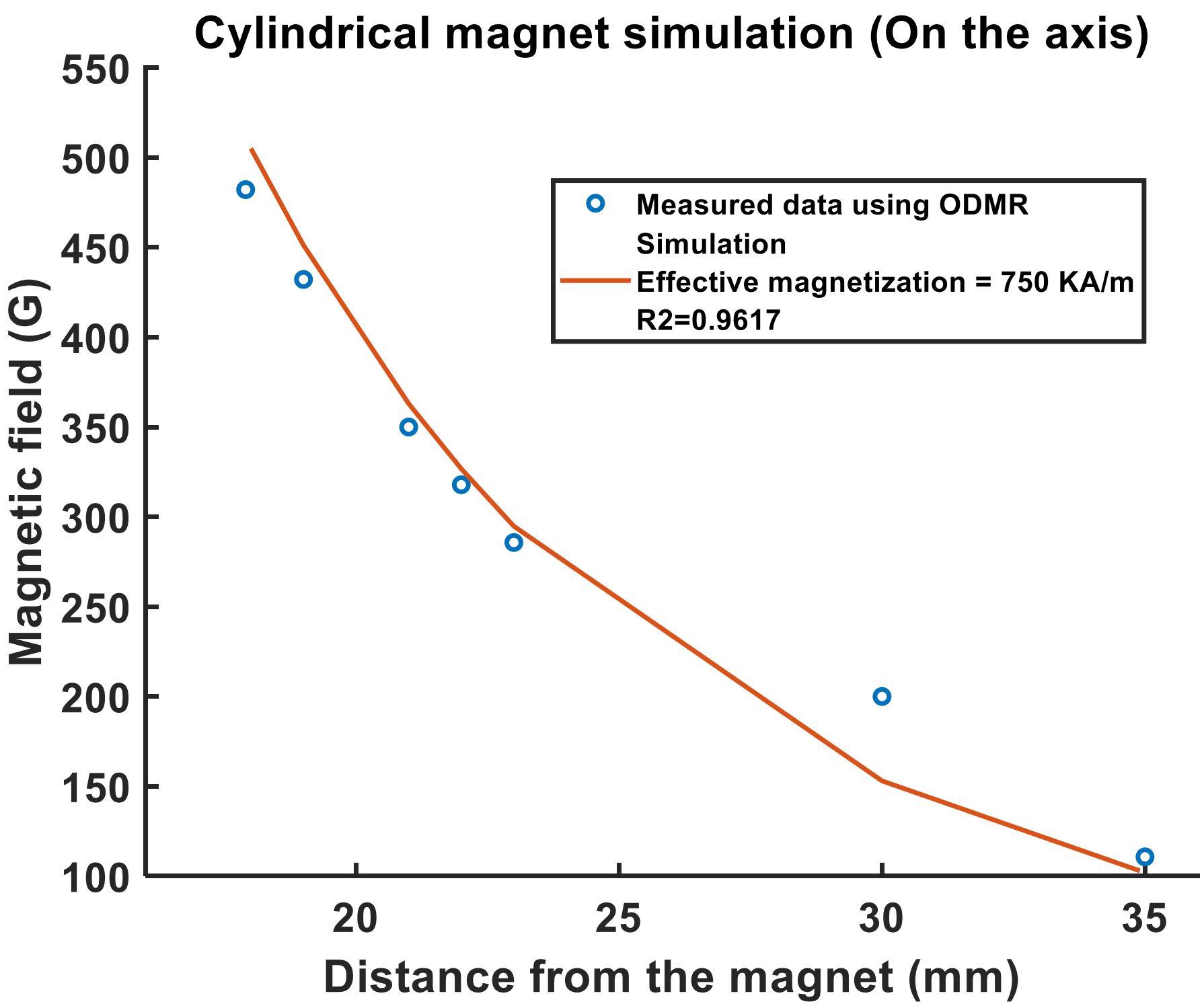

*Figure 2S: Simulated and measured magnetic field on the cylindrical magnet axis.*

# 2. X band EPR for the copper complex

We validated our findings by performing a traditional EPR measurement in our sample.

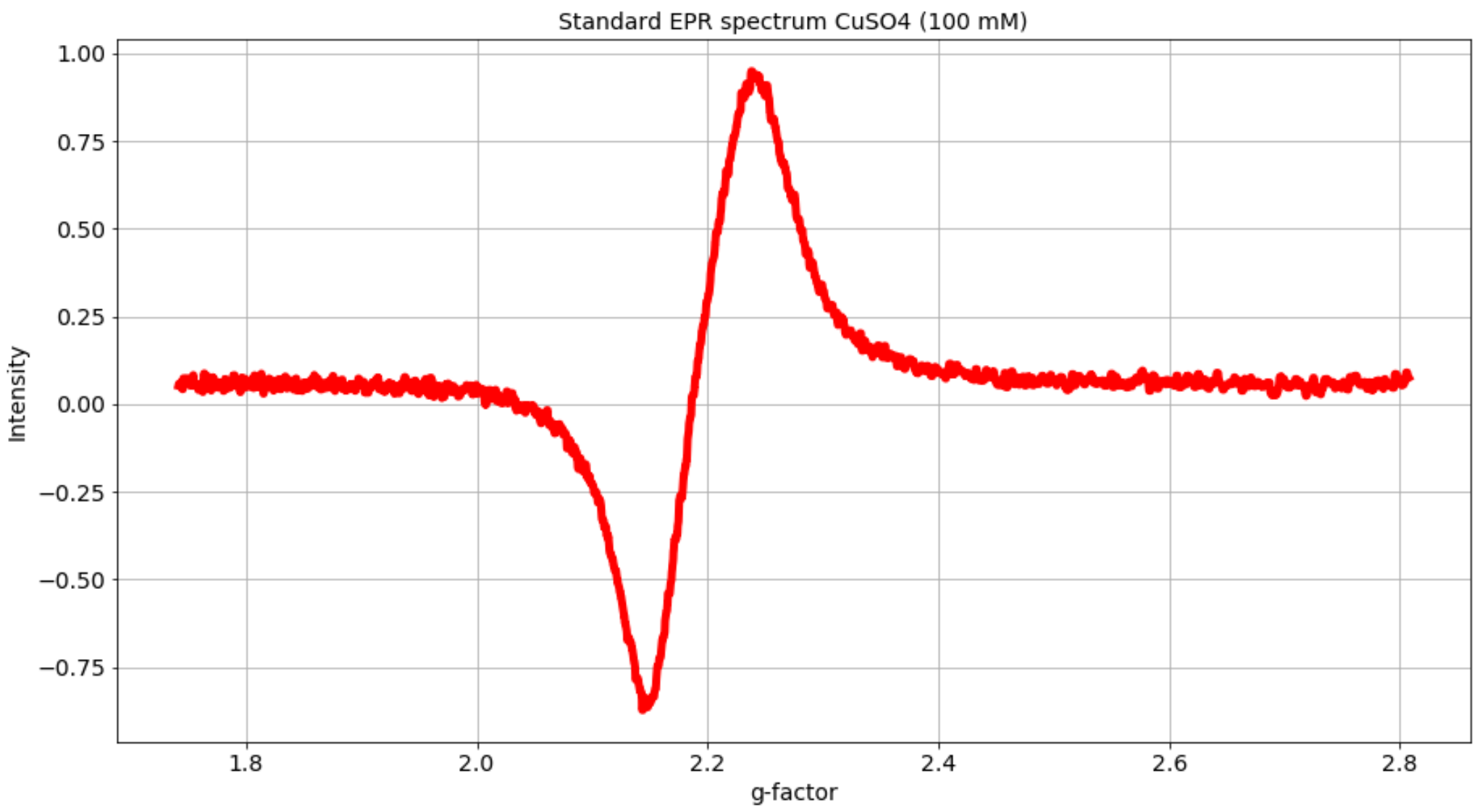

*Figure 2S: EPR spectrum of the $CuSO_4$ sample.*

# 3. Magnetic field visualization

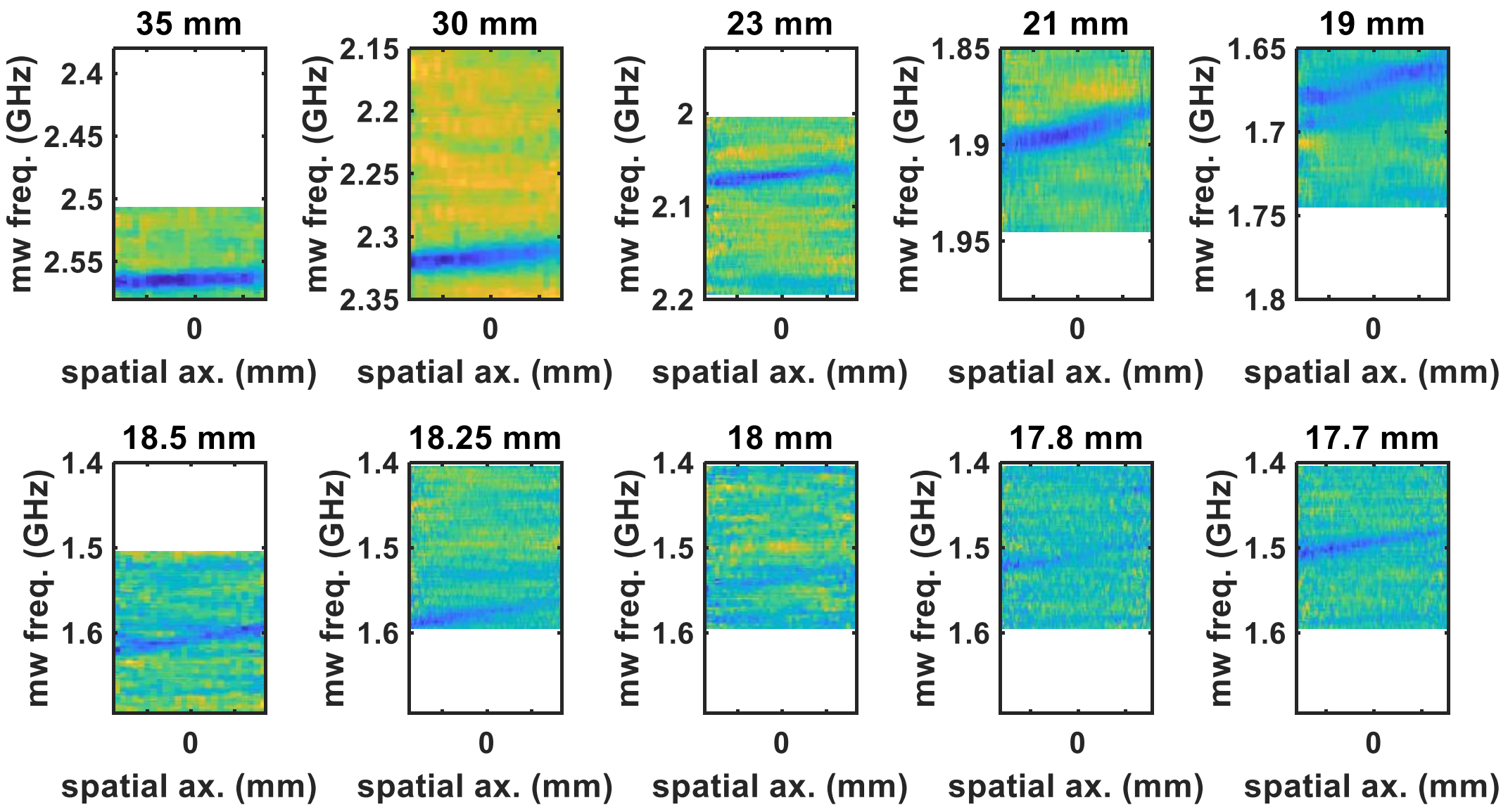

*Figure 3S: Microwave photoluminescence response along the diagonal axis (one pixel width) as a function of the microwave frequency*

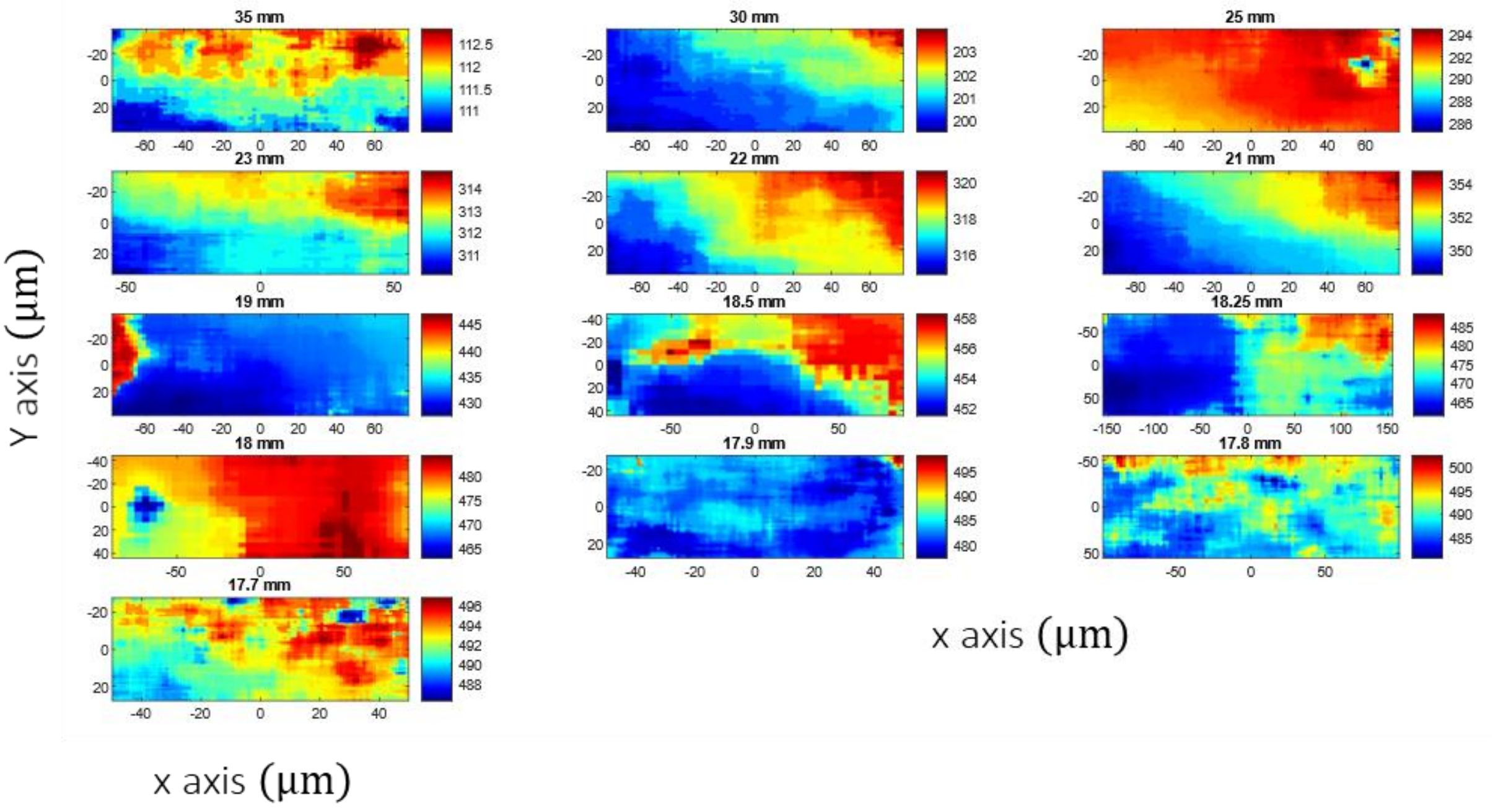

*Figure 4S: Magnetic field maps. The values of the magnetic field results from the Lorentzian fit of the microwave response spectrum. Thus, the magnetic field value is the projection of the local magnetic field on the aligned NV-centre direction.*

# 4. Spin relaxation contrast

We also present the average over the whole relaxometry images as a function of the magnet position.

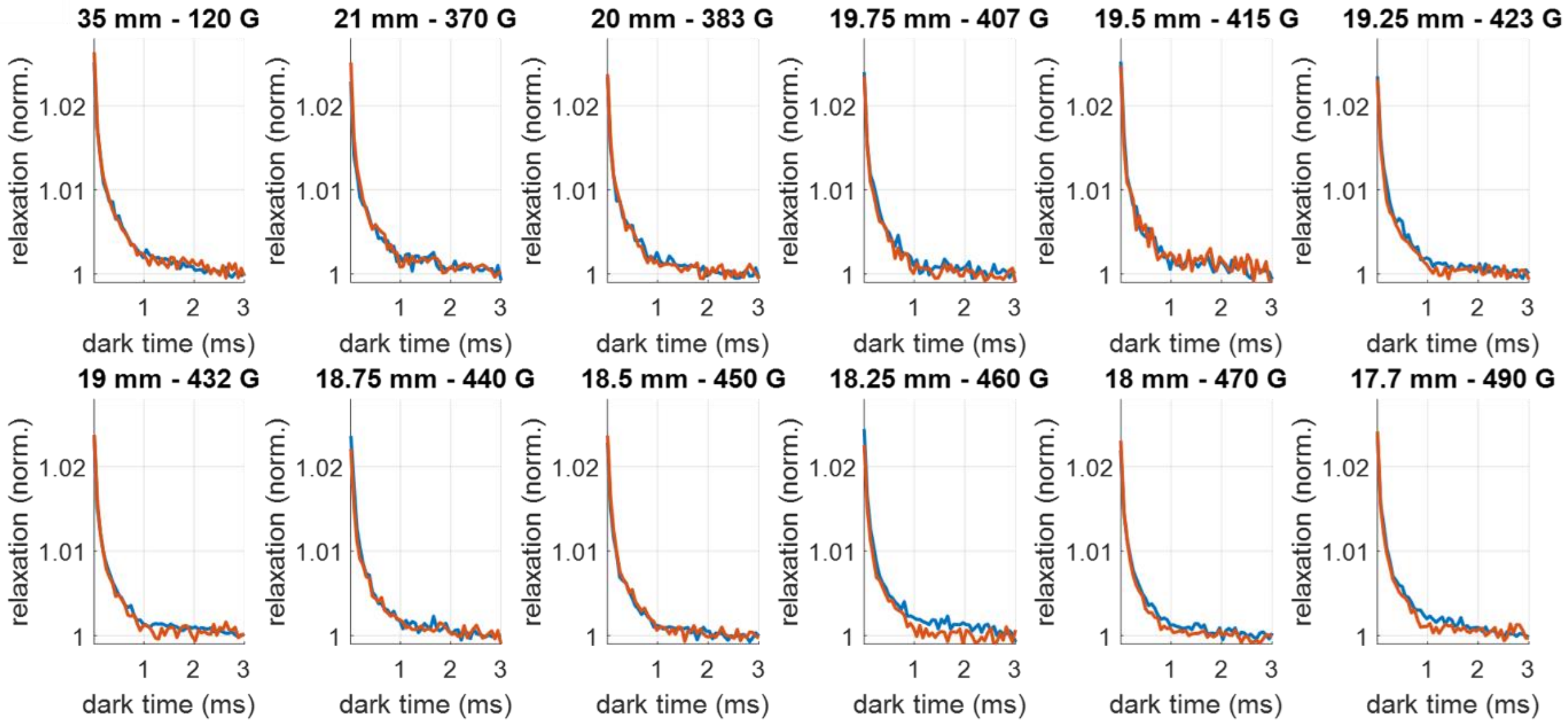

*Figure 5S: Microwave-free electron paramagnetic resonance spectroscopy. Surface average T1-relaxation curves in nitric acid solution (1 mM), blue curves, and nitric acid-copper sulfate solution (100 mM CuSO4 dissolved in 1 mM HNO3), red curves at decreasing distances magnet/diamond, i.e. increasing magnetic fields (A). T1-relaxation curves off and on resonance: 35 mm-110 G (left) and 18.1 mm-460 G (right).*

# 5. Subsampled acquisitions

EPR spectra reconstructed from small subset of the entire acquisition. Chosing only two darktimes ($\tau_s = 1$ and $\tau_l = 3$ ms such that $T_\tau = 4$ ms) but keeping all the sequences $N_{seq} = 10$, or even considering only the first sequence ($N_{seq} = 1$). This diminishes the acquisition times $T_{ac}$ from 31 minutes to 1.6 min and 10 s for each position of the magnet. Figure 6 shows the resulting spectra corresponding to the three situations.

We note that the resonant peak at ~460 G is present in all three spectra. Yet, as expected, we observe a cleaner spectrum when using the complete set of darktimes.

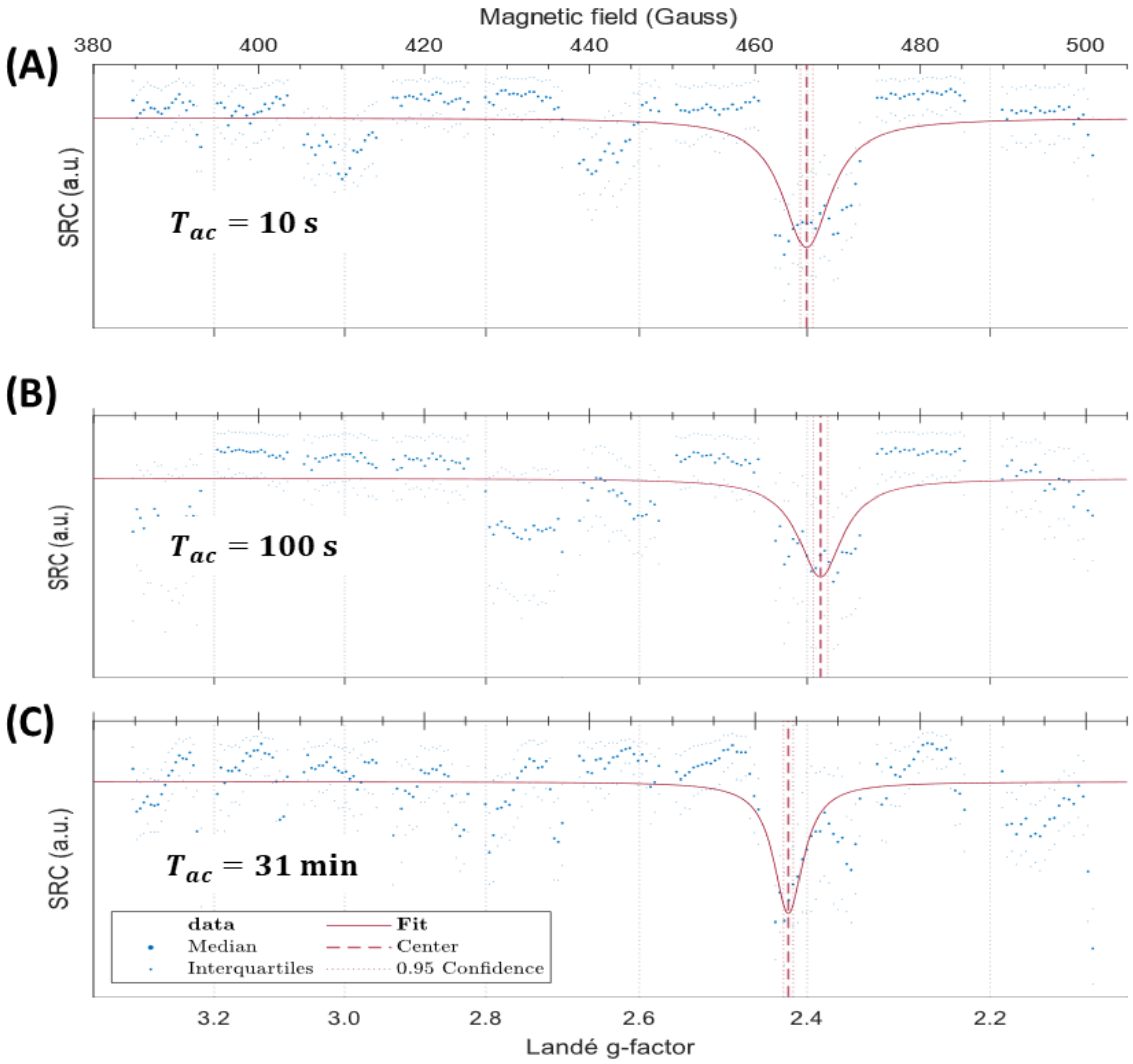

*Figure 6S: Electron paramagnetic spectroscopy using subset of the acquired data. (A) EPR spectrum calculated using PL images at two dark times (1 and 3 ms) $T_\tau = 4$ms and only one single repeat ($N_{seq} = 1$), (B) using PL images at two dark times (1 and 3 ms) and $N_{seq} = 10$, and (C) using the entire dataset. (50 dark times for a total of $T_{tau} = 75$ ms in total) and$N_{seq} = 10$. Estimated acquisition times are indicated for a single position of the magnet. (The shown range corresponds to 11 different position out of the 13 acquired.*

# 6. Apparent coherence time

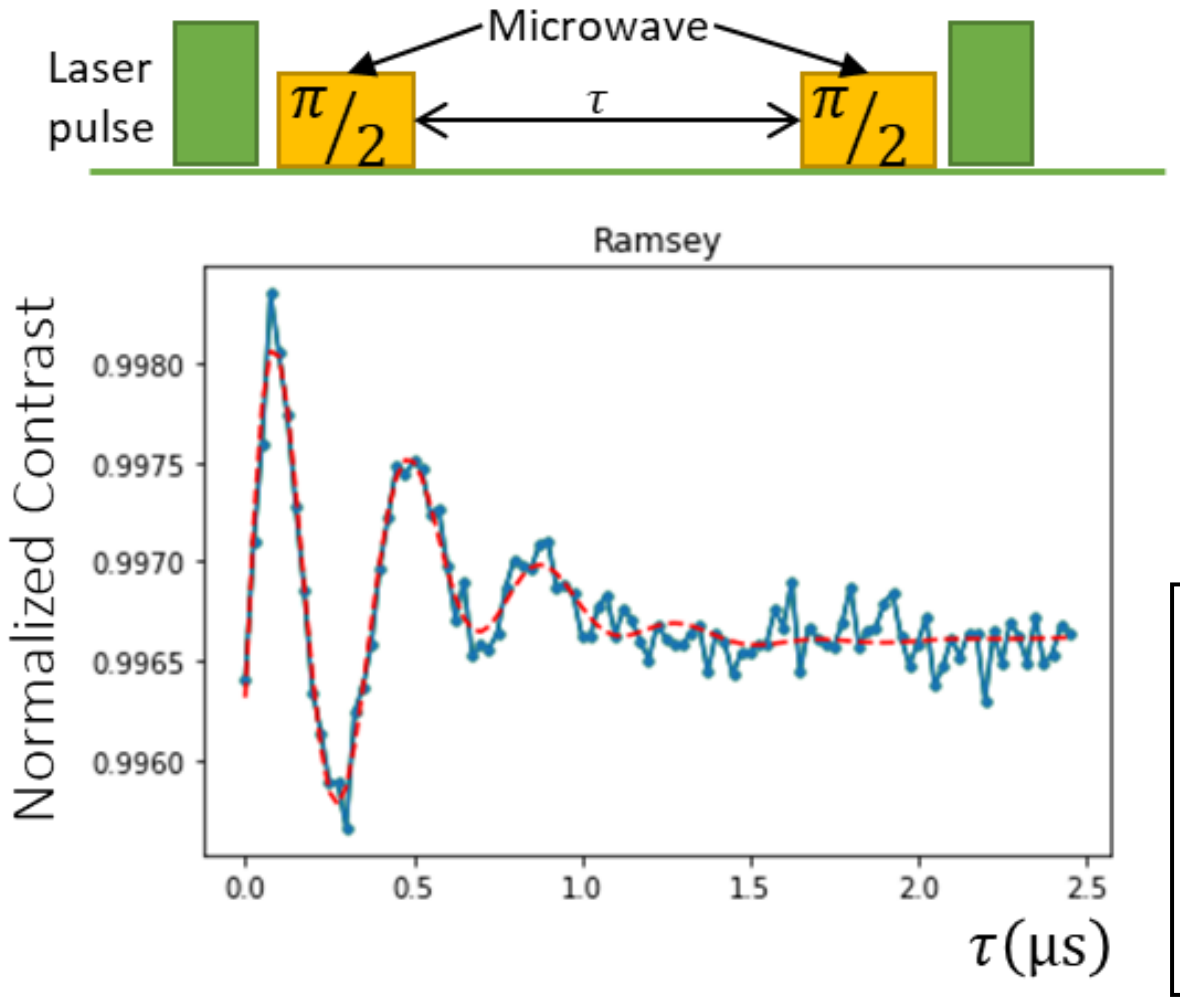

$$Y = C_1 \sin(f_1 \tau) \exp\left(-\frac{\tau}{\tau_1}\right) + C_2 \sin(f_2 \tau) \exp\left(-\frac{\tau}{\tau_2}\right)$$

$C_1 = -2 \cdot 10^{-3}, f_1 = 0.42\ \text{MHz}, \tau_1 = 0.440\ \mu\text{s}$

$C_2 = 2 \cdot 10^{-3}, f_2 = 2.45\ \text{MHz}, \tau_2 = 0.330\ \mu\text{s}$

*Figure 7S: Ramsey fringes at 1.75GHz*

# 7. References

43 Caciagli, A., Baars, R. J., Philipse, A. P., & Kuipers, B. W. (2018). Exact expression for the magnetic field of a finite cylinder with arbitrary uniform magnetization. Journal of Magnetism and Magnetic Materials, 456, 423-432.